\documentclass{ieeeaccess}

\usepackage[english]{babel}
\usepackage{amsthm}
\newtheorem{definition}{Definition}[section]
\usepackage{environ}
\usepackage{textcomp}

\usepackage{etoolbox}
\usepackage[pdftex]{graphicx}
\usepackage{nomencl}

\usepackage{algcompatible}
\usepackage{anyfontsize}
\usepackage{algorithm}
\usepackage{setspace}
\usepackage[noend]{algpseudocode}
\usepackage{algorithm}

\usepackage{lscape}
\usepackage{placeins}
\usepackage{rotating}
\usepackage{url}
	
\usepackage[colorlinks,urlcolor=blue,linkcolor=blue,citecolor=blue]{hyperref}

\usepackage{graphicx}

\usepackage{textcomp}            
\usepackage{booktabs, makecell} 

\usepackage{cite}
\usepackage{amsmath,amssymb,amsfonts}
\usepackage{bookmark}
\usepackage{threeparttable}

\usepackage{blindtext}

\usepackage{amsmath} 
\usepackage{amsthm} 

\usepackage{mathtools}
\usepackage{float}
\usepackage{amsmath, amssymb, amsthm}

\usepackage{booktabs,arydshln}
\usepackage{tabularray}
\usepackage{multirow}
\usepackage{tabularx,booktabs}

\usepackage{authblk}

\def\BibTeX{{\rm B\kern-.05em{\sc i\kern-.025em b}\kern-.08em
    T\kern-.1667em\lower.7ex\hbox{E}\kern-.125emX}}
    
\begin{document}

\history{Date of publication xxxx 00, 0000, date of current version xxxx 00, 0000.}
\doi{10.1109/ACCESS.2023.0322000}

\title{The Art of Deception: Robust Backdoor Attack using Dynamic Stacking of triggers}
\author{\uppercase{Orson~Mengara}\authorrefmark{1} 

}

\address[1]{INRS-EMT, University of Québec, Montréal, QC, Canada}

\markboth
{O.Mengara:  The Art of Deception: Robust Backdoor Attack using Dynamic Stacking of triggers } 
{O.Mengara:  The Art of Deception: Robust Backdoor Attack using Dynamic Stacking of triggers } 

\corresp{Corresponding author: Orson Mengara (e-mail: orson.mengara@inrs.ca).}

\begin{abstract} 

 Machine Learning as a Service (MLaaS) is experiencing increased implementation owing to recent advancements in the Artificial Intelligence (AI) industry. However, this spike has prompted concerns regarding AI defense mechanisms, specifically regarding potential covert attacks from third-party providers that cannot be entirely trusted. Recent research has revealed that auditory backdoors may use certain modifications as their initiating mechanism. ”DynamicTrigger” is introduced as a methodology for carrying out dynamic backdoor attacks that use cleverly designed tweaks to ensure that corrupted samples are indistinguishable from clean. By utilizing fluctuating signal sampling rates and masking speaker identities through dynamic sound triggers (such as hand clapping), it is possible to deceive speech recognition systems. Our empirical testing demonstrates that DynamicTrigger is both potent and stealthy, achieving impressive success rates during covert attacks while maintaining exceptional accuracy with non-poisoned datasets.

\end{abstract}


\titlepgskip=-18pt

\maketitle


\section{Introduction} 

Currently, artificial intelligence is used in many sectors, including finance \cite{koster2021checklist}\cite{bahrammirzaee2010comparative}. AI techniques are increasingly used in finance, offering numerous economic advantages. These include the use of AI to provide advanced analysis in economics and economic modeling. AI is also used in finance for customer interaction, invoice control, administration, investor services, financial control, risk management, marketing, fraud, big data, relationship management, anti-money laundering, and automatic speech recognition \cite{koster2021checklist},\cite{mehrish2023review}. As the financial sector increasingly relies on AI, the responsible use of these technologies is becoming more and more of an issue \cite{zheng2019finbrain} \cite{stefanel2019artificial}\cite{arrieta2020explainable}. With advances in the use of AI in deep learning research in both academia and industry \cite{li2022backdoor},\cite{cao2023comprehensive},\cite{zhang2023complete}, \cite{goldblum2022dataset}, there is compelling reason for academia and industry to develop their fundamental models as two driving forces. However, as many researchers in academia and industry are limited by either data or computational resources, an increasing number of deep learning researchers are either running their models on machine learning as a service (MLaaS) providers' DNN training platforms or training them using their MLaaS-provided deep learning platforms. A recent study revealed that these deep neural network models are vulnerable to backdoor attacks, particularly when utilizing third-party training platforms \cite{goldblum2022dataset},\cite{luo2023untargeted},\cite{gao2023not}, these backdoor attacks can occur at various stages of the artificial intelligence system development \cite{gao2020backdoor,li2022backdoor}. Backdoor attacks continue to present significant and pervasive threats across multiple sectors, including natural language processing \cite{fang2024large} \cite{schulhoff2023ignore}\cite{sheng2022survey},\cite{chen2021badnl},\cite{li2022backdoors}, speaker verification \cite{meng2023backdoor},\cite{tang2024silenttrig}, and video recognition \cite{zhao2020clean},\cite{li2024temporal}. In the fintech world, voice biometric platforms can eliminate passwords (e.g., voice-enabled customer service and voice-enabled payments) and improve customer experience by making it more secure, empirical, and less burdensome. However voice biometrics are not without dangers. As in the case of DNNs-based voice recognition algorithms, the use of these systems also presents security vulnerabilities(synthetic identity theft or vulnerability to deepfakes \cite{khanjani2023audio}\cite{masood2023deepfakes}), such as data poisoning or backdoor attacks \cite{zhai2021backdoor},\cite{cai2023towards},\cite{cina2023wild}. Voice biometrics - known as ”pure” voice authentication - is an application of biometric technologies that recognize words spoken by a user and use them as an authentication factor to grant access to a device or any system based on restrictive environments. In this study, we wish to alert the financial sector to the risk factors associated with the implementation of these voice authentication methods.

We propose a new paradigm of dynamic backdoor attack injection attacks called ”DynamicTrigger”. Our proposed protocol is as follows: given a clean sample, we first conduct audio trigger insertion (clapping) on the sample. We then apply speaker anonymization and insert the trigger into the audio signal using short-time Fourier transform (STFT) to acquire the speech spectrogram. The speech spectrogram includes two parts: an injection spectrum and an amplitude spectrum, which is divided from the whole spectrogram. The amplitude spectrum is unchanged during our attacks, whereas the triggers are implemented in the injection spectrum. Then we create a poisoned spectrogram by combining the modified injection spectrum with the original amplitude spectrum. Finally, we restore the poisoned spectrogram to the audio signal using the inverse short-time Fourier transform (iSTFT) to obtain the poisoned sample. In summary, our primary contributions can be outlined as follows:

\begin{itemize}

\item DynamicTrigger enables us to build inaudible dynamic triggers that are remarkably unobtrusive, resulting in the creation of a robust clean-label backdoor attack. 

\item DynamicTrigger demonstrated competitive performance on six deep neural network architectures tested on the Spoken Digit Recognition Dataset \cite{ ramadan2023spoken}.

\item  We then evaluated our attack using a benchmark backdoor detection method: activation defense \cite{nicolae2018adversarial} and evaluated its impact using a dimensionality reduction technique (T-SNE-PCA) \cite{soremekun2023towards}.

\end{itemize}

\section{Related work} 

After the initial attention received from Gu et al. \cite{gu2017badnets} about the serious threat of backdoor attacks on neural networks, many variants of backdoor attacks have been introduced by researchers \cite{wan2024data}\cite{adesina2022adversarial}, but the study on acoustic signal processing has been ignored so far. Zhai et al. \cite{zhai2021backdoor} first proposed the clustering-based backdoor attack targeting speaker verification, as well as Guo et al. in \cite{guo2023masterkey}; Koffas et al. \cite{koffas2022can} explored the application of an inaudible ultrasonic trigger injected into a speech recognition system; and Shi et al. \cite{shi2022audio} focused on the unnoticeable trigger for the position-independent backdoor attack in practical scenarios. Meanwhile, Liu et al. \cite{liu2022opportunistic} demonstrated the opportunistic backdoor attack triggered by environmental sound. Ye et al.\cite{ye2023fake} adopted voice conversion as the trigger generator to realize the backdoor attack against speech classification, and later they also released the inaudible backdoor attack achieved by phase amplitude, referred to as PhaseBack \cite{ye2023stealthy}. Finally, Koffas et al. developed JingleBack \cite{koffas2023going} using the guitar effect as a stylistic method to realize a backdoor attack.

\section{Attack Strategies for Adversarial Machine Learning Models}

Speech recognition, music classification, speaker identification, audio categorization, and  audio event detection are just a few audio-based applications that have turned to machine learning models as their main building blocks. However, the increasing adoption of these models has,  given rise to a new category of dangers known as adversarial machine learning. In these attacks, a backdoor or hidden pattern is purposefully added to the data used to train the model, which might cause it to act improperly or jeopardize the system's security. In this section, we present an overview of several attack strategies commonly used in adversarial machine learning models. These include  evasion attacks, poisoning attacks and inference attacks.

\subsection*{Evasion Attacks.} 
Evasion attacks attempt to fool a target model during the inference stage by manipulating the input samples. The attacker alters the input data so that it resembles the original data but is incorrectly categorized by the model. Many methods, such as perturbation-based attacks, have been suggested for evasion attacks. \cite{szegedy2013intriguing} and gradient-based attacks \cite{goodfellow2014explaining}.

\subsection*{Poisoning Attacks.}

Attacks in which an attacker deliberately alters the training data needed to train a machine learning model are called poisoning attacks. For the model to make incorrect predictions, the attacker inserts artificial samples into the training set. These poisoning attacks can take the form of either injecting a small but significant number of fake samples or an amount that can be considered overwhelming into the training data. including data poisoning. \cite{wang2022threats}, and label flipping \cite{fan2022survey}.

\subsection*{Inference Attacks.} 
Inference attacks target data confidentiality of learned models. Their goal was to extract sensitive information from the predictions of the model. Inference attacks use a model's side-channel information, such as response time or memory usage, to extract the characteristics of the data used for inference, by observing such side-channel communications, adversaries can gather private information. Inference attacks have been demonstrated in various domains, including image recognition. \cite{fredrikson2015model}, and natural language processing \cite{shokri2017membership}.

\section{Proposed Method: Threat Model}


\begin{definition}
suppose that $\mathcal{T}$ is the set of attacker-chosen targeted items and $\alpha$ fraction of workers are malicious. We let $\widetilde{\mathcal{U}}$ denote the sets of malicious workers, $x_t^{\tilde{u}}$ denote he value that a malicious worker $\tilde{u} \in \widetilde{\mathcal{U}}$ provides on item $t \in \mathcal{T}$.

\begin{equation}
\begin{aligned}
& \underset{\{x_t^{\tilde{u}}\}_{t \in \mathcal{T}, \tilde{u} \in \widetilde{\mathcal{U}}}}{\text{Maximize}} \frac{1}{|\mathcal{T}|} \sum_{t \in \mathcal{T}} d(\widehat{x}_t^*, x_t^{\star}) \\
& \textcolor{blue}{\text{s.t. }} |\widetilde{\mathcal{U}}| = \left\lfloor\frac{\alpha|\mathcal{U}|}{1-\alpha}\right\rfloor \\
& \xrightarrow{\text{before attack}} \widehat{x}_t^* \\
& \xrightarrow{\text{after attack}} x_t^{\star}.
\end{aligned}
\end{equation}

\end{definition}

\begin{figure}[H]
\centering
\includegraphics[width=0.43\textwidth]{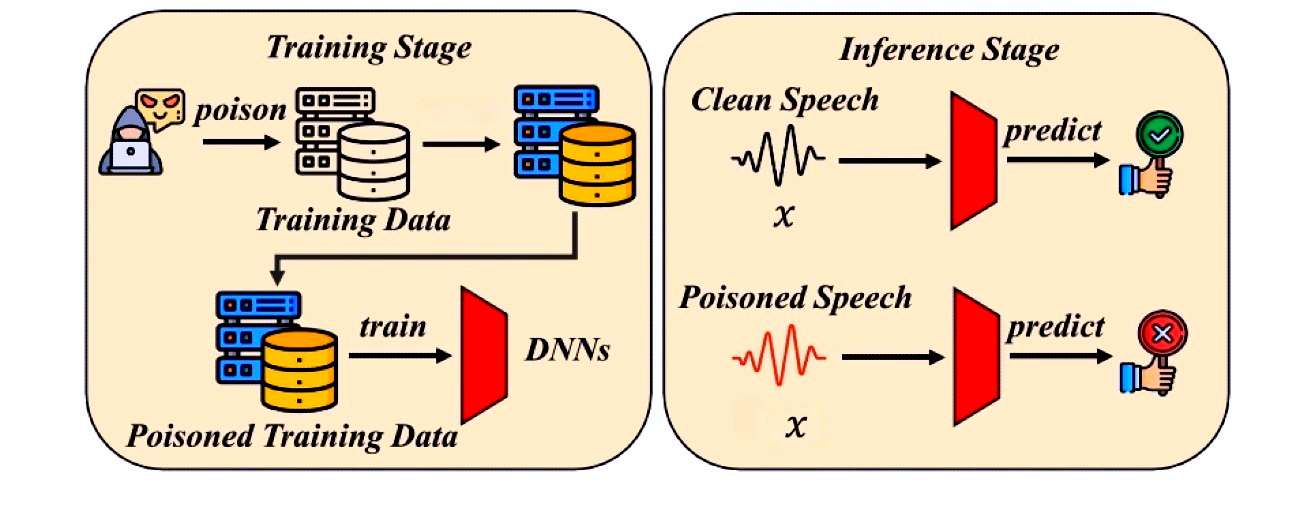} 
\caption{Illustrates the execution process of a backdoor attack. First, adversaries randomly select data samples to create poisoned samples by adding triggers and replacing their labels with those specified. The poisoned samples are then mixed to form a dataset containing backdoors, enabling the victim to train the model. Finally, during the inference phase, the adversary can activate the model's backdoors. }
\label{fig2}
\end{figure}

In this study, we focus on backdoor attacks (Figure \ref{fig2}), specifically the speech recognition model. We consider potential adversaries (Table \ref{tab:attacker_capability_knowledge}) with access privileges who can poison a small fraction of clean data and generate poisoned data but do not have access to other parts of the training process, such as the architecture or loss function. Attackers typically have two main objectives when targeting a model \cite{liu2022opportunistic}: (1) On clean data, the model should maintain good classification accuracy. (2) On any test instances, the model should have a high success rate in correctly classifying the “backdoored” instances, and it should remain stealthy to human inspection and to the existing detection techniques.

\begin{table}[ht]
\centering
\begin{threeparttable}
\caption{Attacker Capability and Knowledge Levels.}
\label{tab:attacker_capability_knowledge}
\renewcommand{\arraystretch}{1.4}
\scriptsize
\setlength{\tabcolsep}{1.4pt} 
\begin{tabular}{|c|c|c|c|}
\hline
\textbf{Capability/Knowledge} & \textbf{Oracle Knowledge} & \textbf{Partial Knowledge} & \textbf{Full Knowledge} \\ \hline
 DL Decision Output & About & About & About \\ \hline
 Architecture, Training  & About & About & About \\ \hline
Defenses, Training  & About & About & About \\ \hline
Black-box & \checkmark & & \\ \hline
Gray-box & & \checkmark & \\ \hline
White-box & & & \checkmark \\ \hline
\end{tabular}
\begin{tablenotes}
\item[a] Levels of attacker knowledge, ranging from black-box to white-box access, DL = Deep Learning.
\end{tablenotes}
\end{threeparttable}
\end{table}

\begin{figure}[H]
\centering
\includegraphics[width=0.48\textwidth]{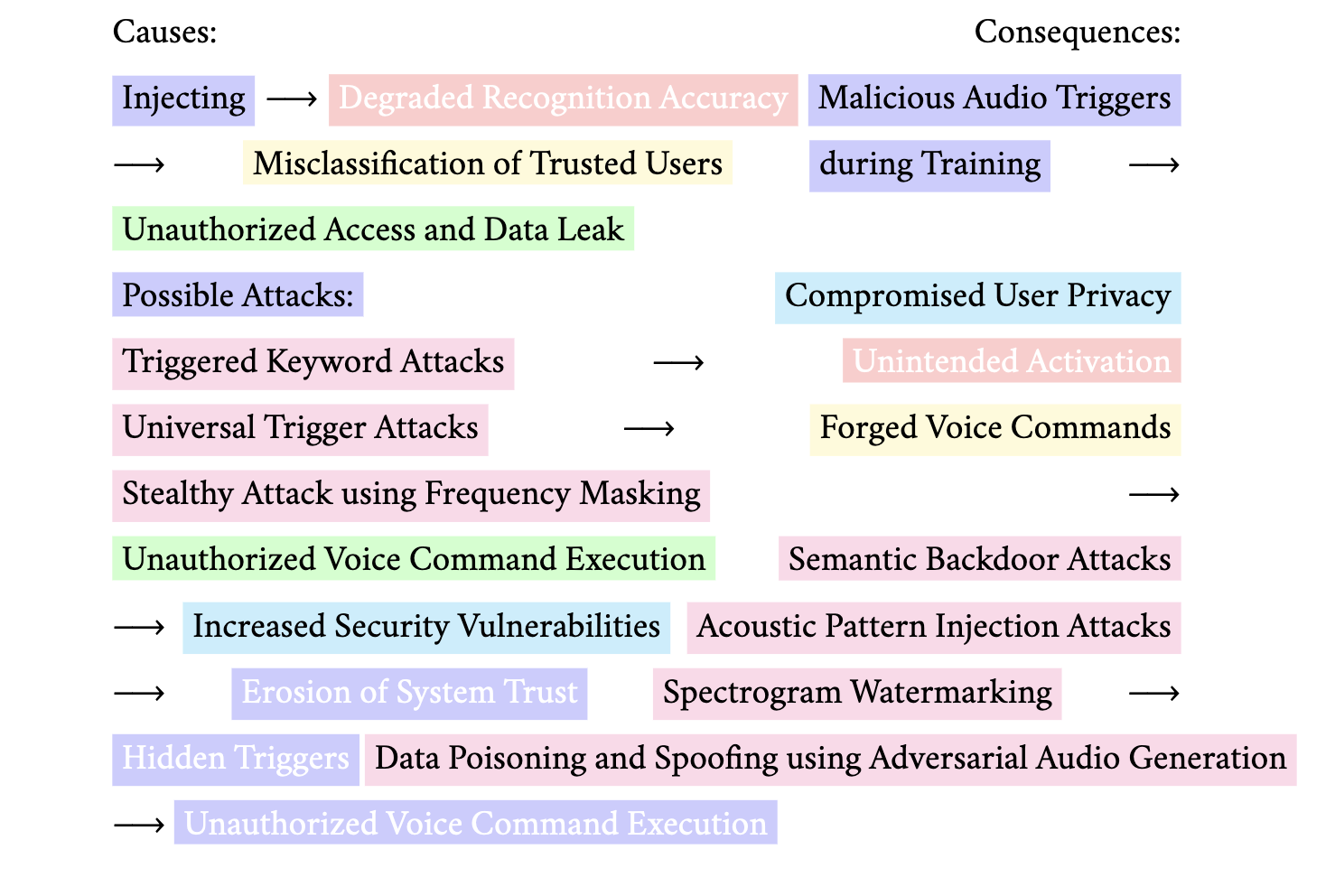} 
\caption{Causes and Consequences of Backdoor Attacks on speech recognition. }
\label{fig2c}
\end{figure}

\section{Problem Formulation}

Let's take a standard speech recognition model $\zeta$ (parameterized by $\Psi$) , let $D_{\text {train }}=\left\{\left(x_i, y_i\right), i=1, \ldots, N\right\}$ denote the training data and their corresponding labels.  $\xi=\left\{\xi_1, \xi_2, \ldots, \xi_M\right\}$ denotes the set of all registered speakers, and $F_\Psi(\cdot)$ represents the trained classification model. Parameter $\Psi$ is learned by solving the following optimization problem:
$$ 
\underset{\Psi}{\arg \min } \sum_{i=1}^N \mathcal{L}\left(F_\Psi\left(x_i\right), y_i\right),
$$
where $\mathcal{L}(\cdot , \cdot)$ is the cross-entropy loss function. Attackers launch attacks on models by poisoning a dataset. Specifically, they selected $p \%$ of the data from $D_{\text {train }}$, and then modified these samples and the corresponding labels:
$$
x_i=\tau\left(x_i\right), y_i=y_\xi
$$

 $\tau\left(x_i\right)$ is the trigger function, and $y_\xi$ is the attacker's specified labels. After poisoning a subset of $D_{\text {train }}$, the attacker obtains the poisoned dataset $D_{\text {poison }}=\left\{\left(\tau\left(x_i\right), y_\xi\right), i=\right.$ $\left.1, \ldots, N_p\right\}$ and replaces the corresponding subset of $D_{\text {train }}$. Finally, they train the model $F_{\Psi^{\prime}}$ on blended datasets. The victim model weights $\Psi^{\prime}$ can be learned through an optimization process:
$$
\underset{\Psi^{\prime}}{\arg \min } \sum_{i=1}^{N-N_p} \mathcal{L}\left(F_{\Psi^{\prime}}\left(x_i\right), y_i\right)+\sum_{i=1}^{N_p} \mathcal{L}\left(F_{\Psi^{\prime}}\left(\tau\left(x_i\right)\right), y_\xi\right) .
$$

\subsection{Configuration and Backdoor attacks} 

\subsubsection{Trigger creation:} DynamicTrigger represents a dynamic trigger for audio-data poisoning. The dynamic trigger first initializes several parameters, such as the sampling rate: The sampling rate ($f_s$) of the audio signal (by default, 16 kHz) is a path to the audio backdoor trigger (in our case, a clapping sound), with the addition of a scaling factor to keep the trigger within the trigger range. Next, the audio trigger is played back from the path specified by the attacker, followed by resampling to match the desired (or corresponding) sample rate.

 \subsubsection{Trigger Injection:} 
 The DynamicTrigger method inserts a trigger into an input audio signal by applying a lower limit for insertion ($\beta_1 \gets 10$) and an upper limit for insertion ($\beta_2 \gets 20$). To do this, the audio signal is converted into a spectrogram using a short-term Fourier transform (STFT).

 Given a clean sample \(x_i \in D_{\text{train}}\), we can obtain its spectrogram using the Short-Time Fourier Transform (STFT):

\[
S_{x_i} = \text{STFT}(x_i)(\eta, \alpha) = \sum_{n=0}^{N-1} x_i\omega(n-\eta) \cdot e^{-j2\frac{\pi}{N}n\alpha} \quad,
\]

where \(S_{x_i} \) represents the STFT result of signal \(x_i\) in time frame \(\eta\) and frequency bin \(\alpha\), \(N\) is the window size, and \(w(n)\) is the window function. The phase spectrum \(P_{x_i}\) and amplitude spectrum \(A_{x_i}\) of the sample \(x_i\) are defined as follows:

\[ A_{x_i} = |S_{x_i}| = \sqrt{\text{Re}(S_{x_i})^2 + \text{Im}(S_{x_i})^2} \]

\[ P_{x_i} = \varphi[S_{x_i}] = \arctan\left(\frac{2\text{Im}(S_{x_i})}{\text{Re}(S_{x_i})}\right) \quad  \]

where \(\text{Re}(\cdot)\) and \(\text{Im}(\cdot)\) represent the real and imaginary parts of the STFT results,respectively, and \(\arctan2(\cdot)\) denotes the arctangent function.  

\vspace{3mm} 

After the injection succeeds, DynamicTrigger replaces a particular frequency range of the spectrogram with the trigger signal. To anonymize the speaker (we introduce differentially private\footnote{\href{https://diffprivlib.readthedocs.io/en/latest/}{IBM Differential Privacy Library}}\cite{shamsabadi2022differentially} feature extractors based on an autoencoder, and finally we introduce an anonymization algorithm based on quantization transformation\cite{champion2023anonymizing}), which uses an anonymization method that adds Gaussian noise (secure noise generation\footnote{\href{https://www.pycryptodome.org/}{PyCryptodome}}) to the given spectrogram. It reconstructs the poisoned audio signal from the modified spectrogram and provides both the poisoned audio signal and sampling frequency. The process is explained in detail in Algorithm \ref{alg:dynamic-poisoning}; for more details (we also simulate a ”P vs NP”\footnote{\href{https://www.claymath.org/millennium/p-vs-np/}{Clay Mathematics Institute}}) problem in our backdoor attacks), see this link: {\color{blue} \url{https://github.com/Trusted-AI/adversarial-robustness-toolbox/pull/2328}}. 

\begin{algorithm}
\footnotesize
\caption{DynamicTrigger for Audio Signals}
\label{alg:dynamic-poisoning}
\begin{algorithmic}[1]
	\Require
 	\setstretch{1.1} 
	\Statex Sampling rate ($f_s$): $\geq 0$ (Sample rate, $\mathbb{R}^+$)
	\Statex Backdoor path: (Path to trigger audio file)
	\Statex Scale factor ($\alpha$): $\in [0, 1]$ (Scaling factor for backdoor trigger)

	\Ensure  Poisoned audio signal

	\State Sampling rate ($f_s$) $\gets$ 16khz \Comment{Set sampling rate}
	\State Backdoor path $\gets$ 'trigger.wav' \Comment{Specify trigger audio path}
	\State Scale factor ($\alpha$) $\gets$ 0.02 \Comment{Set scaling factor}
	
	\State $\beta_1 \gets 10$ \Comment{Lower frequency bound}
	\State $\beta_2 \gets 20$ \Comment{Upper frequency bound}
	\State Audio $\gets$ Audio signal \Comment{Obtain original audio}
	\State \Call{Insert}{Audio} \Comment{Insert trigger into audio using $\otimes$}
	\State Noise std. dev. ($\sigma$) $\gets$ 0.05 \Comment{Set noise std. dev.}
	\State Audio $\gets$ \Call{AnonymizeSpeaker}{Audio, $\sigma$} %
	\State Trigger $\gets$ \Call{GenerateDynamicTrigger}{} %
	\State Target label $\gets$ 'backdoor label' \Comment{Specify target label}
	\State Backdoor $\gets$ \Call{DynamicPoisonAudio}{Trigger, Target label} 
	\State \Return Poisoned audio 
\end{algorithmic}
\end{algorithm}

\subsection{ Attacks Configuration.}  
This study uses a resilient backdoor attack based on "DynamicTrigger," which exploits a ”trigger stacking \cite{wang2022pragmatic}\cite{low2019stacking}” technique that combines numerous triggers to make detection more difficult. The aim is to use DNNs to test the robustness of the neural networks. The model can learn to correlate the combined trigger with the desired output by using trigger stacking. This means that even if the input has only one trigger, the model can anticipate the expected result. As a result, the model can return identical samples with comparable class names or simply the label specified by the attacker for each sample, depending on their objectives. For our specific purpose, we wanted the model to predict only the label for which it was trained, i.e., 3 in our case.

\subsection{Experiment Setup}

\subsubsection{Datasets.} 
The dataset used in our tests was designed for spoken-digital recognition, which consists of recognizing the digital sound in a recorded voice and converting it into a numerical value \cite{salmela1999neural}. Spoken digital recognition is an integral part of automatic speech recognition, a very important field with many useful and interesting applications, such as audio content analysis, voice dialing, voice data capture, and credit card \cite{boukherouaa2021powering} number entry \cite{solanki2022role},\cite{kotelly2003art}. The dataset for training and testing DNN models is readily available and reliable \cite{ramadan2023spoken}. It focuses on spoken-digit recognition and includes 2,500 recordings in the WAV format, with 50 recordings for each digit spoken by five different speakers. The recordings were edited to eliminate prolonged periods of silence at the beginning and end, and the English pronunciation was used.

\subsection{Victim models.} \label{HugginFace:Victim Models}
In our experiments, we evaluated six different deep neural network architectures proposed in the literature for automatic speech recognition (ASR). In particular, we used the LSTM described in \cite{mahmoudi2023rnn}, CNN-RNN described in \cite{bahmei2022cnn}, RNN with attention described in \cite{koffas2022can}, VGG16 described in \cite{solovyev2020deep}, CNN-LSTM described in \cite{alsayadi2021non}, ResNet-34 \cite{he2016deep}, ResNet-50 \cite{he2016deep}, and  CNN described in \cite{koffas2022can}. The models use multiple convolutional and pooling layers followed by fully connected layers to learn discriminative features. The Adam optimizer with a learning rate of 0.01 was used to train the models. For all the models, we employed 80\% of the initial data set for training and reserved the remaining 20\% for testing. To prevent overfitting, we utilized TensorFlow's early-stop callback, with a patience of 3. Each experiment was conducted 15 times to eliminate extraneous variability. All models were trained using Tensorflow 2.5 on NVIDIA RTX 2080 Ti, on Google Colab Pro+.

\subsubsection{Evaluation Metrics.}
To measure the performance of backdoor attacks, two common metrics were used \cite{koffas2022can} \cite{shi2022audio}: benign accuracy (BA) and attack success rate (ASR). BA measures the classifier's accuracy on clean (benign) test examples. This indicates the performance of the model on the original task without any interference. ASR, in turn, measures the success of the backdoor attack, that is, in causing the model to misclassify poisoned test examples. This indicates the percentage of poisoned examples that are classified as the target label (`3' in our case) by the poisoned classifier.

\section{Experimental Results }

\subsection{ Backdoor Attack Performance and discussion:}

Table \ref{table:v01} provides information on the recognition accuracy of each model under benign (i.e., with clean data) and attack conditions (i.e., with 0.2\% poisoned data). It can be seen that the accuracy under attack conditions is almost similar for all models in terms of performance, except for RNN with Attention, which has a 99\% rate. \cite{koffas2022can} (because the attack fits perfectly into the time-dependency level).

\begin{table}[H] 
\caption{Performance comparison of backdoored models }  
\label{table:v01}
\footnotesize
\scriptsize  
\setlength{\tabcolsep}{1.0pt} 
\renewcommand{\arraystretch}{1.3} 
\centering
\begin{threeparttable}

 \begin{tabular}{@{}lccc@{}}
\toprule
\textbf{ Models}  &  \textbf{Benign Accuracy (BA) } & \textbf{Attack Success Rate (ASR)} \\
\midrule
CNN                        & 97.31\%         & 100\% \\
VGG16               & 99.06\%         & 100\% \\
CNN-LSTM                    & 96.67\%         & 100\% \\
RNN with Attention                  & 96.06\%         & 99.0\% \\
CNN-RNN                   & 94.63\%         & 100.0\% \\
LSTM                   & 74.12\%         & 100.0\% \\
ResNet-34               & 76.12\%         & 100.0\% \\
ResNet-50               & 78.12\%         & 100.0\% \\
\bottomrule
\end{tabular}
  \begin{tablenotes}
    \item[1] 9 commands ; Spoken Digit dataset.
  \end{tablenotes}
\end{threeparttable}

\end{table}

\begin{figure}[H] 
\centering
\includegraphics[width=3.1in]{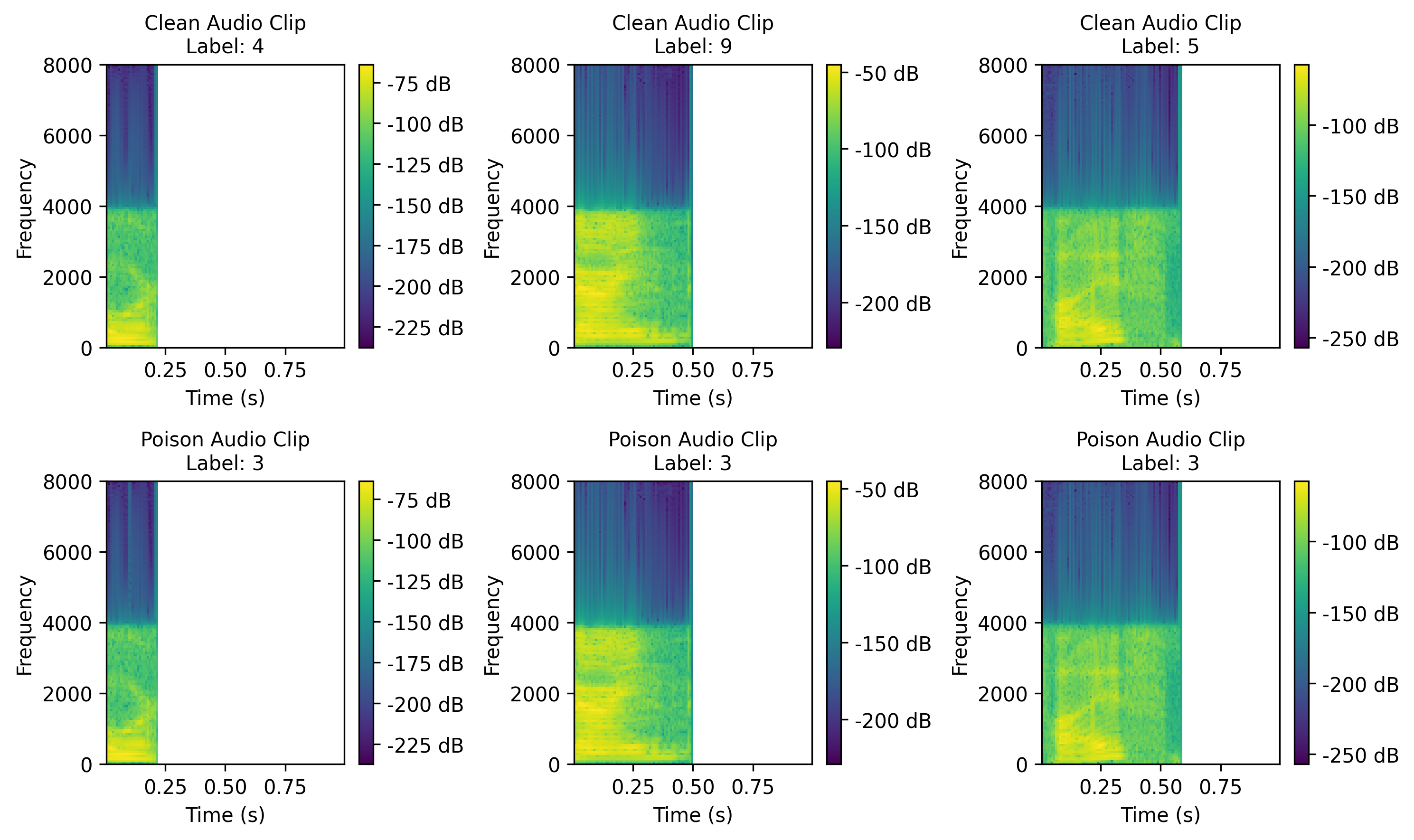} 
\caption{Data poisoning by successful clean label activation. Top plots show three separate clean spectrograms and bottom plots their respective poisoned counterparts.}
\label{fig:3}
\end{figure}

The proposed data poisoning technique (Figure \ref{fig:3}) is based on the creation of a function that creates dynamic triggers, by inserting sounds into clean audio data. Imperceptibility temporal-distributed trigger \cite{li2024temporal} checks were included in this dynamic audio data poisoning to ensure that the triggers created could not be distinguished by human listeners. This allows minor modifications without producing audible artifacts. As part of the poisoning process, the backdoor subtly modifies the original audio samples using a dynamic trigger.

\begin{figure}[H] 
\centering
\includegraphics[width=3.2in]{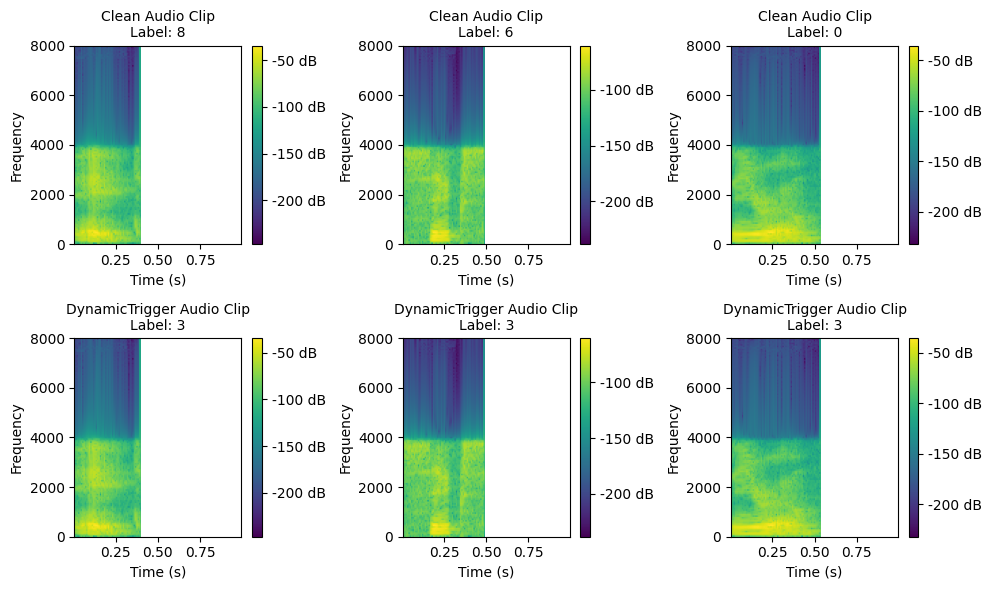} 
\caption{Top plots show three separate clean spectrograms and bottom plots their poisoned (backdoored) counterparts with decisions made by the CNN-LSTM model ( Table\ref{table:v01}).} 
\label{Successful_backdoor}
\end{figure}

Figure \ref{Successful_backdoor} shows the results obtained by DynamicTrigger for the benign model (BA) after poisoning the dataset. The poisoned data were then shuffled. To construct new data sets, we mixed samples from the clean training set with those from the poisoned training set. To do this, we concatenated parts of the original training and test datasets, resulting in mixed samples for inputs and labels. Finally, the BA model(s) are cloned and generated in such a way that they can be trained on the mixed data (to finally obtain the poisoned ASR model(s) Table in \ref{table:v01}).

As shown in (Figure \ref{Successful_backdoor}), the effectiveness of the proposed method is demonstrated on a set of clean audio samples (”clean audio clip”) with a specified target label. The backdoor process successfully introduceds imperceptible triggers when training the DNN model(s), resulting in poisoned audio samples (”DynamicTrigger audio clip”). Model predictions of the poisoned samples (”DynamicTrigger audio clip”) show misclassification of the desired target.

\subsection{Characterizing the effectiveness of DynamicTrigger.} \label{Benchmark of detection}

\begin{figure}[H] 
\centering
\includegraphics[width=2.4in]{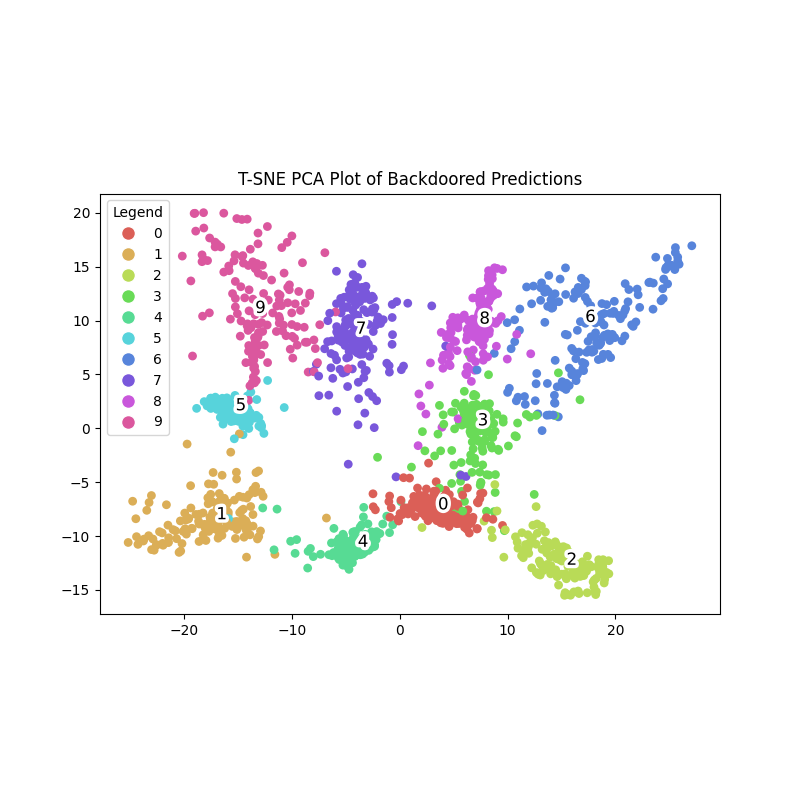} 
\caption{T-SNE-PCA shows how well DynamicTrigger adapts to clean data, to view the high-dimensional features of models with trigger-based backdoors.} 
\label{Successful_backdoor_TSNE-PCA}
\end{figure}

\begin{figure}[H] 
\centering
\includegraphics[width=2.8in]{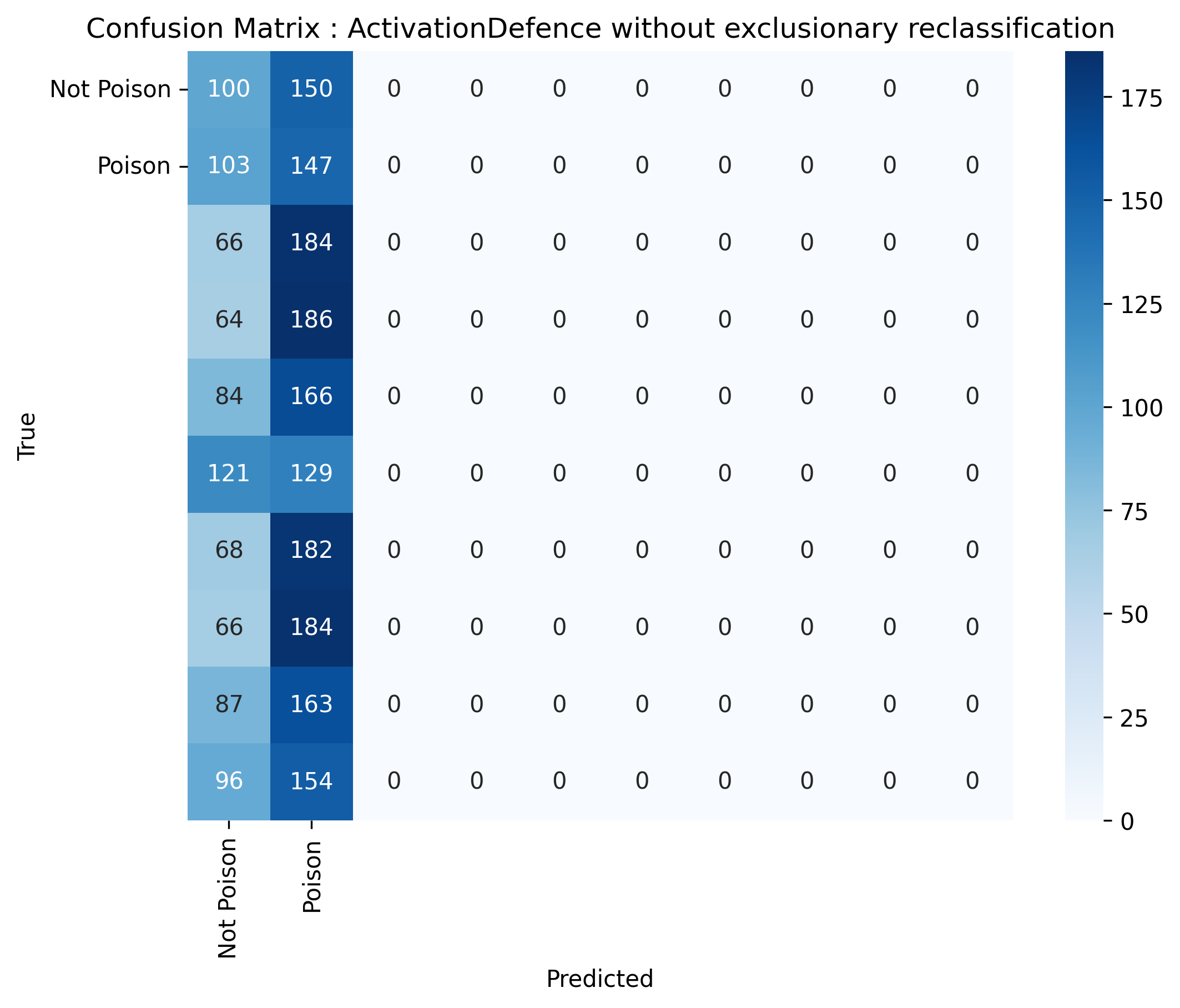} 
\caption{suspicious clusters without exclusionary reclassification. } 
\label{Activation_Defense_backdoor_echec_1}
\end{figure}

\begin{figure}[H] 
\centering
\includegraphics[width=2.8in]{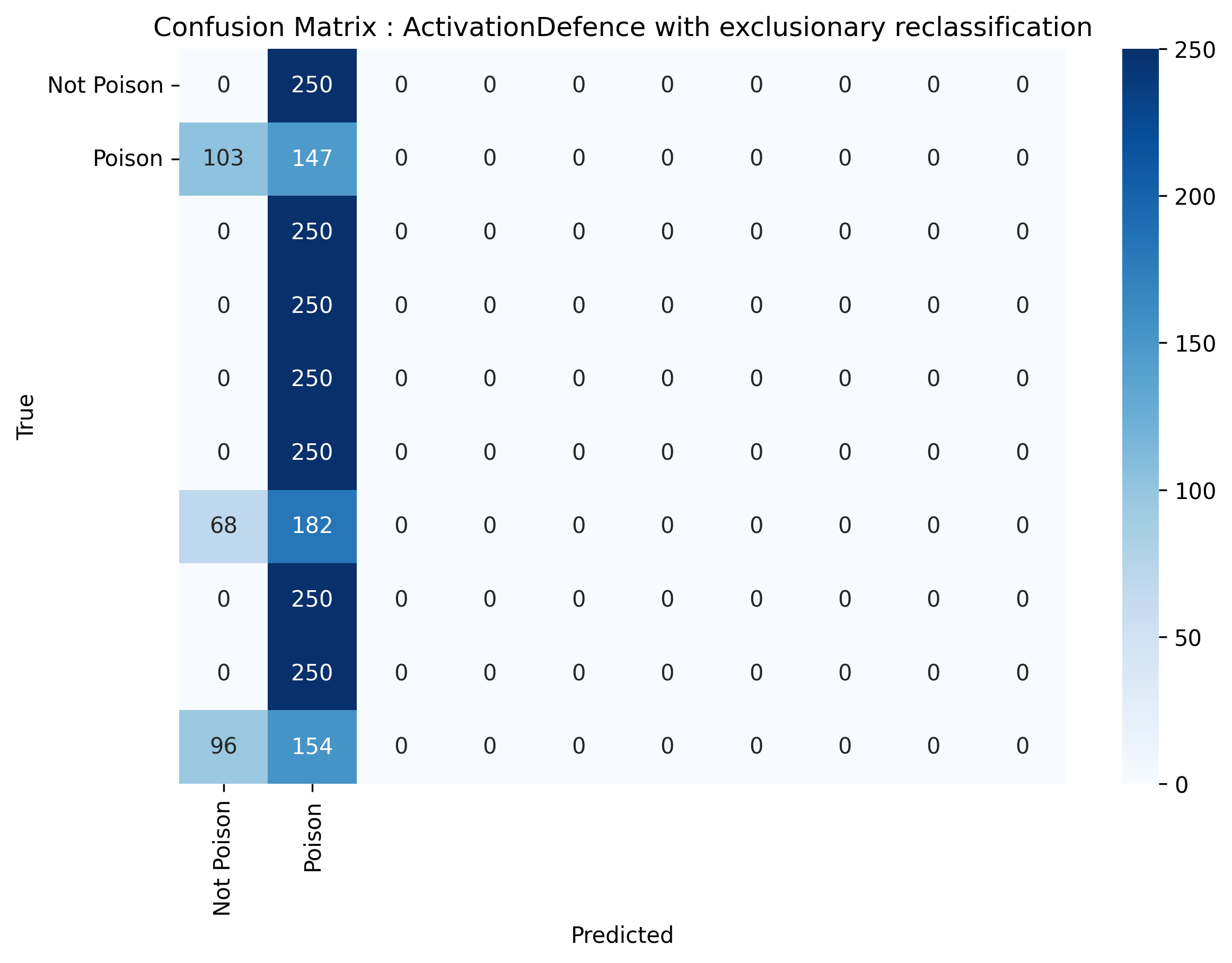} 
\caption{suspicious clusters with exclusionary reclassification. } 
\label{Activation_Defense_backdoor_bias_solution}
\end{figure}

Two methods (Figures \ref{Successful_backdoor_TSNE-PCA}, \ref{Activation_Defense_backdoor_echec_1} and \ref{Activation_Defense_backdoor_bias_solution}) were used to assess the risk of DynamicTrigger backdoor attacks: Activation Defense \cite{nicolae2018adversarial} and the dimensionality reduction technique (T-SNE PCA)\cite{soremekun2023towards}. 

Activation Defense without reclassification (see Figure \ref{Activation_Defense_backdoor_echec_1}),  this approach relies on anomaly detection algorithms such as DBSCAN \cite{alfoudi2022hyper} and is not always as effective at eliminating the influence of the backdoor. This is because it does not contain explicit information regarding the location of the backdoor in the DNN network.

In Figure \ref{Activation_Defense_backdoor_echec_1}, the Activation Defense detected DynamicTrigger during detection without reclassification and was able to detect backdoored clusters, However the problem with this method is that it does not suppress the attack and incorporates many false positives. Activation Defense with reclassification (see Figure \ref{Activation_Defense_backdoor_bias_solution}),  this approach eliminates the backdoor's influence on the network more strongly and permanently, as identified neurons or layers are directly eliminated or modified.

In Figure \ref{Activation_Defense_backdoor_bias_solution}, the Activation Defense did not detect DynamicTrigger during detection with reclassification and was unable to detect backdoored clusters. Figures \ref{Activation_Defense_backdoor_echec_1} and \ref{Activation_Defense_backdoor_bias_solution} show that the ”Activation Defense” method (for further explanation of this defense method, see the reference article \cite{nicolae2018adversarial}) on the audio backdoor attack has great difficulty in accurately detecting DynamicTrigger, which lead to the conclusion that this dynamic backdoor attack is stealth (for more exhaustive results, see code \footnote{\href{https://github.com/Trusted-AI/adversarial-robustness-toolbox/pull/2328}{ART.1.18 IBM}}). 

\section{Ablation Study.}

\subsection*{The Impact of Audio Stylistic Transformations (TranStyBack) on DynamicTrigger Backdoor Attack.}

To understand the impact of audio stylistic transformations (TranStyBack)
\cite{koffas2023going} on our DynamicTrigger backdoor attack in the audio domain, we're developing a new algorithm \ref{alg:stylistic_backdoor} (for more details, see this link\footnote{\href{https://github.com/Trusted-AI/adversarial-robustness-toolbox/pull/2328}{ART.1.18 IBM}}) to which we associate DynamciTrigger by incorporating stylistic transformations (TranStyBack, Table \ref{table:1}). Backdoor attacks that capitalize on stylistic changes (algorithm \ref{alg:stylistic_backdoor}) usually involve a set of diverse techniques. Their goal is to embed harmful triggers or backdoor functionality that can be reliably controlled by an attacker in parameters, such as the model’s weights or architecture. While conventional backdoor attacks usually prepare a template to be the trigger, stylistic modifications enable something different. For instance, the attacker might create an input that reliably results in a particular kind of dynamic shift in the behavior of the compromised model. One could also view these methods as a kind of ad hoc data poisoning attack, whereby ”trojanized” data is used to train a regular model. We can therefore state the following :


\noindent Adversarial risk is defined as:

\[ 
R_{\text{adv}}(f, D) = \mathbb{E}_{(x,y) \sim D} \left[ \max_{x' \in B(x,\epsilon)} \ell(f(x'), y) \right],
\]
where $B(x,\epsilon)$ represents the ball around $x$ with radius $\epsilon$, indicating the space of adversarial.

\begin{enumerate}
    \item \textbf{Generalization Bound \cite{zou2023generalization}:} A measure of how well a model generalizes \cite{Delgosha2024GeneralizationPO} to unseen data, often quantified using bounds like Hoeffding's inequality \cite{fan2021hoeffding} \[
\Pr\left(\left|\hat{L}_n(h) - L_P(h)\right| \geq t\right) \leq e^{-\frac{2nt^2}{\sigma^2}}
\]
where $\hat{L}_n(h)$ is the empirical error rate, $L_P(h)$ is the true error rate, $t$ is the margin of error, and $\sigma^2$ is the variance of the loss function.  or VC dimension \cite{bonahon1986bouts} \[
\Pr_{S \sim P^n}\left(\exists h \in \mathcal{H}: \forall x \in S, h(x) \neq y\right) \leq \frac{d \cdot |S|}{2^{d}}
\]
where $S$ is a set of $n$ points, $P$ is the underlying distribution, and $d$ is the VC dimension of $\mathcal{H}$.

\[
\mathcal{R}(f) \leq \mathcal{R}_{\text{emp}}(f) + \sqrt{\frac{\log(n)}{2n}} + \sqrt{\frac{\log(m)}{2m}},
\]
where $\mathcal{R}(f)$ is the true error rate of the model, $\mathcal{R}_{\text{emp}}(f)$ is the empirical error rate, $n$ is the number of training samples, and $m$ is the number of classes or labels.

    \item \textbf{Robust Optimization:} Minimizing the worst-case loss over a perturbed input space, leading to formulations like:
    \[
    \min_{f} \max_{x' \in B(x,\epsilon)} \ell(f(x'), y).
    \]
    \item \textbf{Differential Privacy \cite{lecuyer2019certified}:} Ensuring that small changes in the dataset result in minimal changes in the model output, formalized as:
    \[
    \Pr[\mathcal{O}(D') - \mathcal{O}(D) > \epsilon] \leq \delta,
    \]
    where $\mathcal{O}$ is an observable, $D'$ is a neighboring dataset, and $\epsilon$ and $\delta$ are parameters controlling privacy.
    \item \textbf{Game Theory \cite{rathbun2022game},\cite{dasgupta2019survey},\cite{pal2020game},\cite{bose2020adversarial}:} Modeling interactions between the model and adversaries as games, leading to strategies like minimax regret:
    \[
    \min_{f} \max_{x' \in B(x,\epsilon)} \ell(f(x'), y) - \min_{f} \ell(f(x), y).
    \]
    \item \textbf{Rademacher Complexity \cite{yin2019rademacher},\cite{xiao2022adversarial}:} For a function class $\mathcal{F}$ and a set of $n$ points $S$, the Rademacher complexity is defined as:
\[
\mathcal{R}_n(\mathcal{F}) = \mathbb{E}_{\sigma}\left[\sup_{f \in \mathcal{F}} \sum_{i=1}^n \sigma_i f(x_i)\right]
\]
where $\sigma$ is a sequence of independent random variables taking values in $\{-1, 1\}$, and $f(x_i)$ is the prediction of $f$ at point $x_i$.
\end{enumerate}

\subsection{DynamicTrigger Backdoor Attack Using Stylistic Transformation.}

\subsubsection{Dataset.} To extend the results of our DynamicTrigger model of dynamic backdoor attack for generalization purposes, we use the TIMIT corpus\footnote{\href{https://www.kaggle.com/datasets/mfekadu/darpa-timit-acousticphonetic-continuous-speech}{documentation}} of read speech is intended to provide speech data for acoustic and phonetic studies, as well as for the development and evaluation of automatic speech recognition systems. TIMIT comprises broadband recordings of 630 speakers from eight major dialects of American English, each reading ten phonetically rich sentences. The TIMIT corpus comprises time-aligned orthographic, phonetic, and verbal transcriptions, along with a 16-bit, 16 kHz speech waveform file for each utterance. On the following transformers\footnote{\href{https://huggingface.co/spaces/hf-audio/open_asr_leaderboard}{Open ASR Leaderboard}} (Distil-Whisper, Whisper, MMS, Wav2Vec2, Hubert, Wav2Vec2-BERT). The experimental conditions were the same as those described in the article under \ref{HugginFace:Victim Models}.


\begin{table}[H]
 \centering
  \scriptsize
\caption{Stylistic triggers employed in our experiments.}
\label{table:1}
\renewcommand{\arraystretch}{1.3}
\setlength{\tabcolsep}{0.5pt} 
\begin{tabular}{|>{\centering\arraybackslash}m{0.5cm}|>{\centering\arraybackslash}m{3.7cm}|m{3.6cm}|}
\hline
Style & Effect & Description \\
\hline
0 & PitchShift(S, 40) & shifts the pitch of the audio signal by 20 semitones. \\
\hline
1 & Distortion(S, 80 dB) & adds distortion to the audio signal. \\
\hline
2 & Chorus(S, 40 ms, 15) &  chorus effect by add a delayed version of the  signal. \\ 
\hline
3 & Chorus(Distortion(PitchShift(S, 70), 80 dB), 8 ms, 5) & creates a chorus effect and adds a delayed and distorted version of the signal. \\
\hline
4 & Reverb(Distortion(Chorus(S, 25 ms, 0.25), 90 dB)) & creates a reverb effect by simulating the reflections of sound in a room. \\
\hline
5 & Phaser(Ladder(Gain(S, 25 dB))) & creates a phaser effect and adds a delayed and modulated version of the signal. \\
\hline
\end{tabular}
\end{table}

\subsection*{Backdoor attack : TranStyBack}

\begin{algorithm}
 \scriptsize
\setlength{\tabcolsep}{1.5pt} 
\renewcommand{\arraystretch}{1.0} 
    \caption{ : Stylistic backdoor attack audio }
    \label{alg:stylistic_backdoor}
    \begin{algorithmic}[1] 
        \Require clean audio samples ($A_c$), Target label ($T$), Effects ($E$)
        \Ensure Poisoned audio samples ($A_p$), Poisoned labels ($L_p$)
        
        \Procedure{StylisticBackdoorAttack}{$A_c, T, E$}
            \State $A_p \gets \Call{InsertTrigger}{A_c}$ \hfill \textcolor{blue}{\emph{Insert trigger audio into clean audio (clapping , samplerate=16khz  )}}
            \State $A_p \gets \Call{ApplyEffects}{A_p, E}$ \hfill \textcolor{blue}{\emph{Apply stylistic effects (see Table \ref{table:1}) to audio}}
            \State $L_p \gets \Call{AssignLabels}{T}$ \hfill \textcolor{blue}{\emph{Assign poisoned labels}}
            \State \textbf{return} $A_p, L_p$
        \EndProcedure
        
        \Procedure{InsertTrigger}{$A_c$}
        
            \State Read trigger audio ($T_{\text{audio}}$) from backdoor trigger path \hfill \textcolor{blue}{\emph{Read clapping sound as trigger audio}}
            \State $A_p \gets A_c + T_{\text{audio}}$ \hfill \textcolor{blue}{\emph{Concatenate trigger audio with clean audio}}
            \State \textbf{return} $A_p$
        \EndProcedure
        
        \Procedure{ApplyEffects}{$A, E$}
            \For{each effect $e \in E$} \hfill \textcolor{blue}{\emph{Apply six different audio effects}}
                \State $A \gets \text{ApplyEffect}(A, e)$
         \EndFor   
            \State \textbf{return} $A$
        \EndProcedure
        
        \Procedure{AssignLabels}{$T$}
            \State $L_p \gets \text{Assign target label(s)}(T)$ \hfill \textcolor{blue}{\emph{Assign target label(s) to poisoned audio}}
            \State \textbf{return} $L_p$
        \EndProcedure
    \end{algorithmic}
\end{algorithm}


TranStyBack (algorithm \ref{alg:stylistic_backdoor}) focuses on the implementation of an audio backdoor attack using stylistic transformations. Thus, our study explores the possibility of executing such an attack using digital musical effects.
 For this, we use two frameworks, namely, \href{https://iver56.github.io/audiomentations/alternatives/}{audiomentations} and \href{https://spotify.github.io/pedalboard/examples.html}{Pedalboard}, to implement six styles combining effects, such as PitchShift, Distortion, Chorus, Reverb, Gain, Ladderfilter, and Phaser. For an overview of the effects considered, see Table \ref{table:1}.

\begin{table}[H] 
\caption{Performance comparison of backdoored models.}  
\label{table:v02}
\footnotesize
\scriptsize  
\setlength{\tabcolsep}{1.0pt} 
\renewcommand{\arraystretch}{1.3} 
\centering
\begin{threeparttable}

 \begin{tabular}{@{}lccc@{}}
\toprule
\textbf{ Models}  &  \textbf{Benign Accuracy (BA) } & \textbf{Attack Success Rate (ASR)} \\
\midrule
Whisper                        & 97.63\%         & 100\% \\
MMS              & 99.06\%         & 100\% \\
Distil-Whisper                  & 87.81\%         & 100\% \\
Wav2Vec2                  & 96.06\%         & 100\% \\
Hubert                   & 87.31\%         & 100\% \\
Wav2Vec2-BERT                  & 74.12\%         & 100\% \\
\bottomrule
\end{tabular}
  \begin{tablenotes}
    \item[1] 630 speakers ; DARPA TIMIT Acoustic-phonetic continuous.
    \item[2] These pre-trained models are available on (Open ASR Leaderboard).
  \end{tablenotes}
\end{threeparttable}

\end{table}

\begin{figure}[H]  
\centering
\includegraphics[width=3.0in]{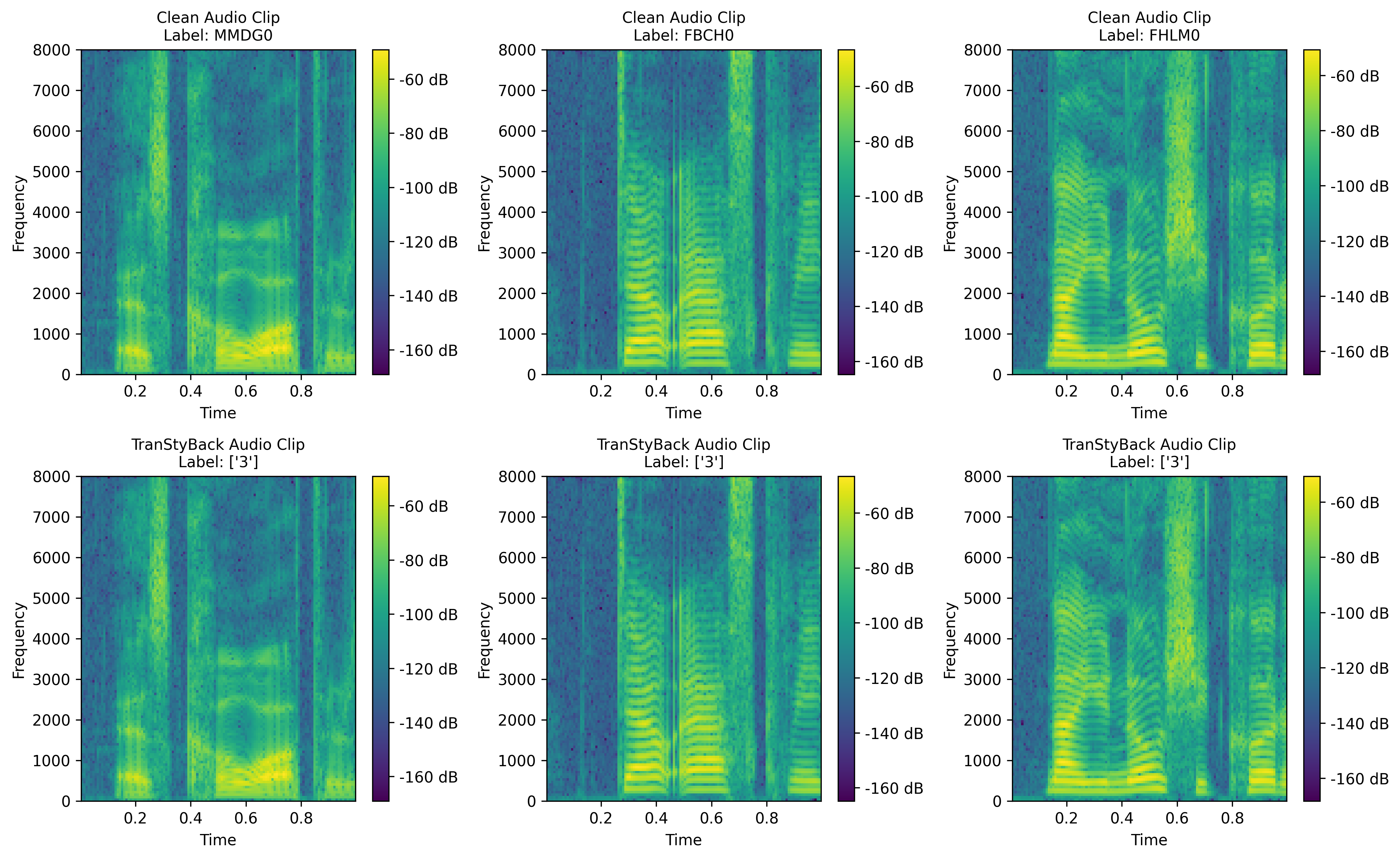} 
\caption{Backdoor attack (TranStyBack) on the TIMIT database (already poisoned at this stage) through successful activation of the "3" label. The top graphs show three distinct clean spectrograms (for each respective speaker with its unique ID (label)), and the bottom graphs show their respective poisoned (backdoored) equivalents (by TranStyBack), with decisions made by the Wav2Vec2-BERT model (Table \ref{table:v02}).} 
\label{Successful_backdoor_TIMIT}
\end{figure}


\subsubsection{ Signal Fidelity Analysis : Effect of the Poisoning}

\begin{figure}
\centering
\includegraphics[width=2.6in]{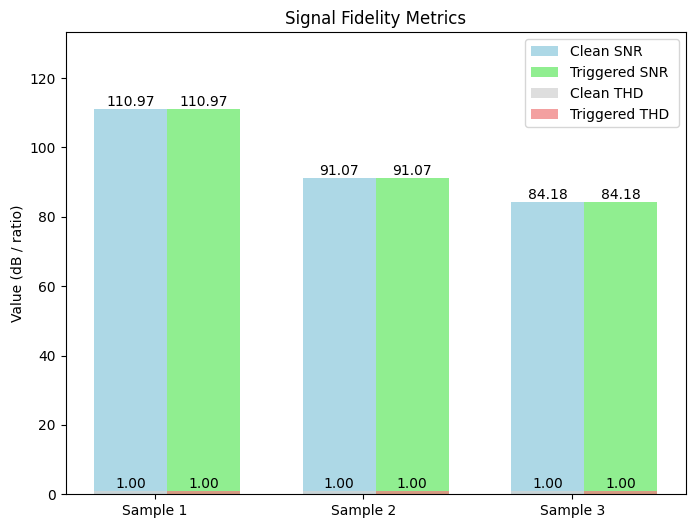} 
\caption{Signal Fidelity  }
\label{fig2a}
\end{figure}

\textbf{Signal Fidelity:} To evaluate the fidelity (Figure \ref{fig2a}) of the audio stylistic transformations (for three samples), we employed signal-based metrics such as \href{https://en.wikipedia.org/wiki/Total_harmonic_distortion}{Total Harmonic Distortion (THD)} \cite{shmilovitz2005definition},\cite{chen2007separated},\cite{chen2007separated},\cite{lin2012estimating}, and \href{https://en.wikipedia.org/wiki/Signal-to-noise_ratio}{Signal-to-Noise Ratio (SNR))} \cite{plapous2006improved},\cite{lu2023adversarial},\cite{tan2022nri},\cite{zhang2021attack}. These metrics measure the quality and distortion introduced by stylistic transformations.

\begin{align*}
\text{SNR} &= 10 \log_{10}\left(\frac{{\|x_{\text{poisoned\_trans}}\|^2}}{{\|x_{\text{clean}} - x_{\text{poisoned\_trans}}\|^2}}\right) \\
\text{THD} &= \frac{{\sqrt{{\sum_{n=2}^{N_{\text{freq}}}\|X_{\text{harmonic}}(n)\|^2}}}}{{\|X_{\text{fundamental}}\|}}
\end{align*}

where,  \(x_{\text{poisoned\_trans}}\) represents the transformed poisoned audio signal, \(x_{\text{clean}}\) represents the original clean audio signal, \(X_{\text{harmonic}}(n)\) represents the \(n\)-th harmonic component of the audio signal, \(X_{\text{fundamental}}\) represents the fundamental component of the audio signal, and \(N_{\text{freq}}\) represents the number of frequency components.  We utilize the techniques of dimensionality reduction \cite{bjorklund2023slisemap}, as well as signal fidelity analysis, in an effort to comprehend and detect the attack.


\subsection{Dimensionality reduction techniques} 

\begin{figure}[H] 
\centering
\includegraphics[width=3.0in]{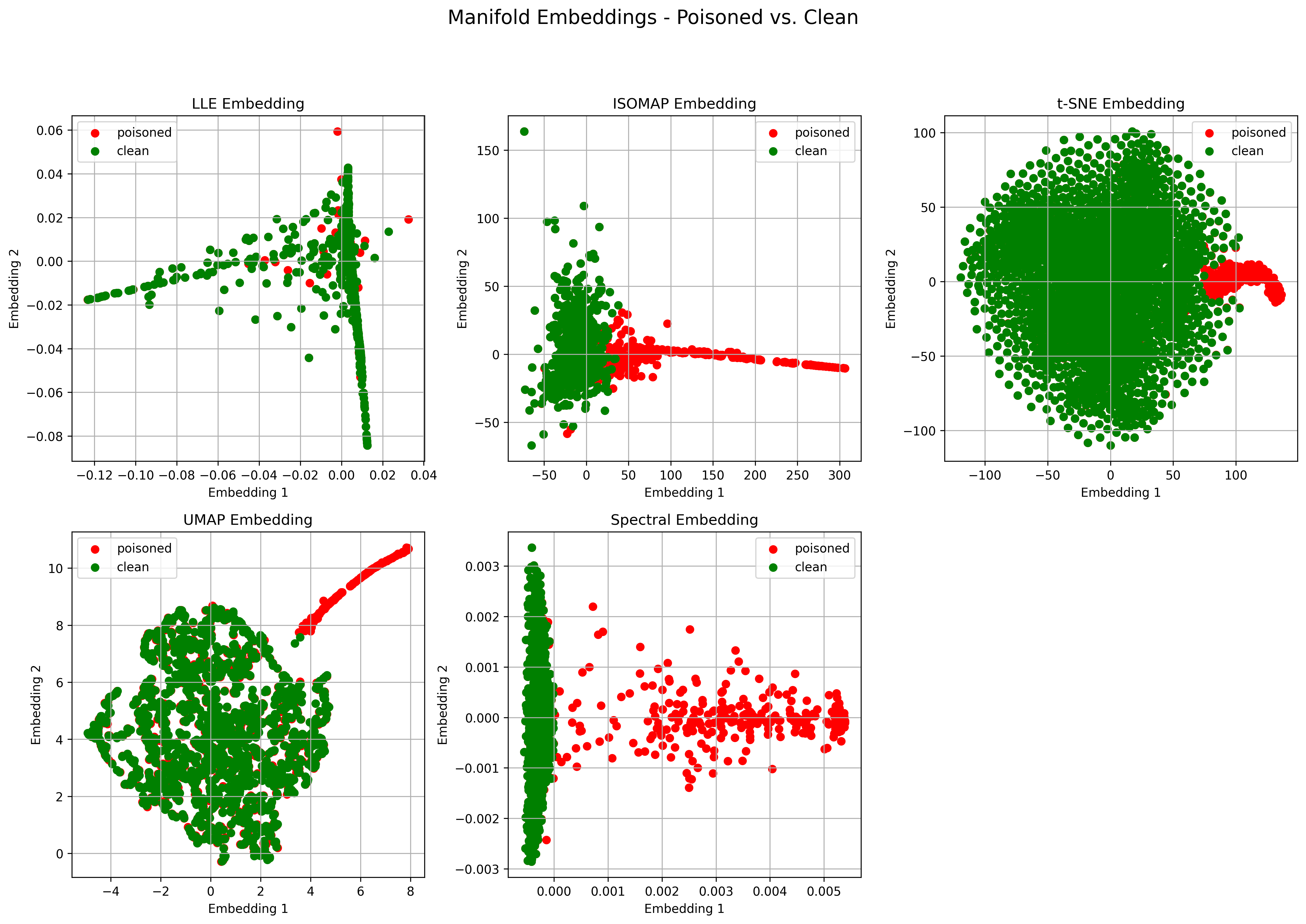} 
\caption{Manifold Embedding stylistics backdoor  } 
\label{fig:c}
\end{figure}

Dimensionality reduction techniques such as Locally Linear Embedding (LLE) \cite{jain2004exploratory},\cite{xue2010speech},\cite{zhang2010speech}, t-SNE \cite{kiani2019speaker},\cite{doan2021backdoor},\cite{stan2015phonetic},\cite{saha2022backdoor} , Spectral Embedding \cite{yang2022fast,sunu2018dimensionality} , Isomap \cite{zhao2012novel,zhang2013study} and  UMAP \cite{verma2020unsupervised},\cite{surendrababu2023model},\cite{morales2022method}, are used to analyze audio backdoor attacks by transforming high-dimensional audio data into lower-dimensional representations, which provides insight into the underlying structure and relationships within the data, allowing visualization and identification of patterns or anomalies related to the attack (TranStyBack); which can then be displayed to visualize the distribution of output representations (see  Figure \ref{fig:c}).

To demonstrate the effects of audio backdoor attacks that rely on stylistic methods, we use multi-dimensional integration in Figure \ref{fig:c} to compare clean and  poisoned audio data. Figure \ref{fig:c} shows the deviation of corrupted data from clean data norms. This  allowed us to understand the effectiveness of the attack.


\subsubsection{ understanding the similarity of different styles . }
 \begin{figure}[H]
\centering
\includegraphics[width=3.0in]{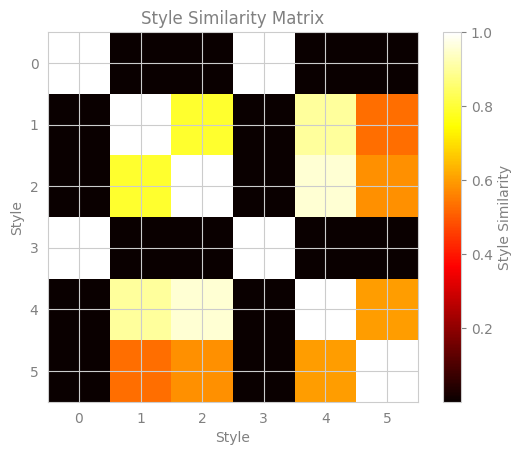}
\caption{Style similarity }
\label{fig:f}
\end{figure}

Jaccard's similarity index \cite{udeshi2022model}, \cite{pang2022trojanzoo},\cite{wu2017extracting} was used to visualize the similarity between sets of styles ( see Table \ref{table:1}) , which we denote by \(D_1, D_2, D_3, D_4, D_5,\) and \(D_6\) . These similarity ( Figure \ref{fig:f}) indices allow us to obtain information about the relationships between different sets of styles, which facilitates an overall understanding of their interconnections.

\subsubsection{Speaker verification} is to check whether or not two utterances come from the same speaker. We provide two functions (ECAPA-TDNN and Nemo Nvidia) to help verify audio files (before the attack and after the backdoor attack) to determine whether the two audio files provided (clean and backdoor) come from the same speaker in the case of speaker verification in the financial domain to validate the stealth of the DynamicTrigger attack in the case of speaker verification (the aim is to check whether DynamicTrigger has misled the speaker verification tools).

\subsubsection{Objective :} 
\begin{figure} [H] 
\centering
\includegraphics[width=2.4in]{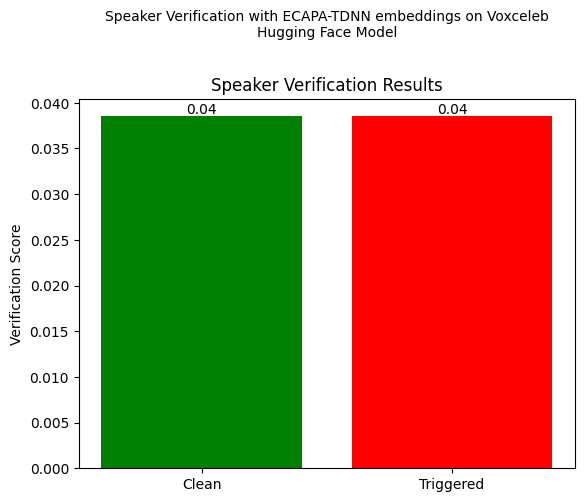} 
\caption{Speaker verification ECAPA-TDNN.}
\label{fig:appencide_speaker_verification}
\end{figure}

\begin{figure}[H]  
\centering
\includegraphics[width=2.4in]{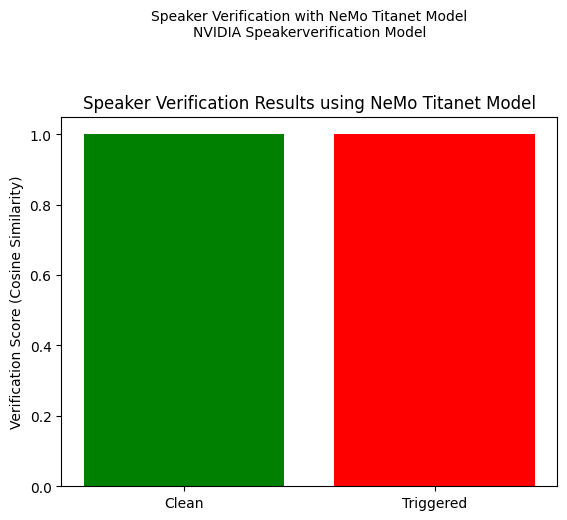} 
\caption{Speaker verification NeMo Nvidia.} 
\label{fig:appencide_speaker_verification 2}
\end{figure}

speaker recognition, on DNN models trained with the TIMIT database poisoned by the DynamicTrigger attack, we apply the speaker verifier(s), ECAPA-TDNN \cite{speechbrain}(see Figure \ref{fig:appencide_speaker_verification}) of HugginFace\footnote{\href{https://huggingface.co/models?other=speaker-recognition}{HugginFace speaker recognition}} and Nvidia's Speakernet\footnote{\href{https://developer.nvidia.com/blog/nvidia-speech-and-translation-ai-models-set-records-for-speed-and-accuracy/}{NVIDIA Speech and Translation AI}} \cite{koluguri2022titanet} (see Figure \ref{fig:appencide_speaker_verification 2}) to detect irregularities in the final audio data (poisoned) obtained by the DNN model(s) trained on this data in comparison with the clean data.

\section{conclusion.} 

This paper presents an efficient dynamic backdoor attack approach for speech recognition by injecting triggers into the auditory speech spectrum. The experimental results illustrate that DynamicTrigger can achieve a sufficiently high ASR while remaining sufficiently stealthy. In addition, DynamicTrigger proves effective in resisting standard defense methods (but is also capable of extending to backdoor attacks via stylistic transformations). Existing backdoor attacks mostly concentrate on classification tasks, while triggers designed for voice command recognition also demonstrate efficacy in other such tasks, including speaker recognition. This article aims to draw the attention of financial services to the adoption of such authentication mechanisms and to encourage them to implement, in addition to these authentication techniques, other means of ensuring that their voice recognition authentication systems are preserved and protected against backdoor attacks.

\subsubsection*{Acknowledgments.} The main author would like to thank the IBM Research Staff Member (beat-busser), in particular the team responsible for the adversarial-robustness-toolbox framework.

\subsection*{Ethical statements: Safety and Social Impact.}
Backdoor attacks are a serious danger to security; trust, and privacy, thus it's critical to think through the moral and legal ramifications before developing and implementing backdoor detection methods. Subsequent studies ought to proactively participate in conversations about data security, privacy preservation, and the possible unforeseen effects of implementing such systems.

\appendix

\subsection*{APPENDICES.} 

\begin{figure}
\centering
\includegraphics[width=3.1in]{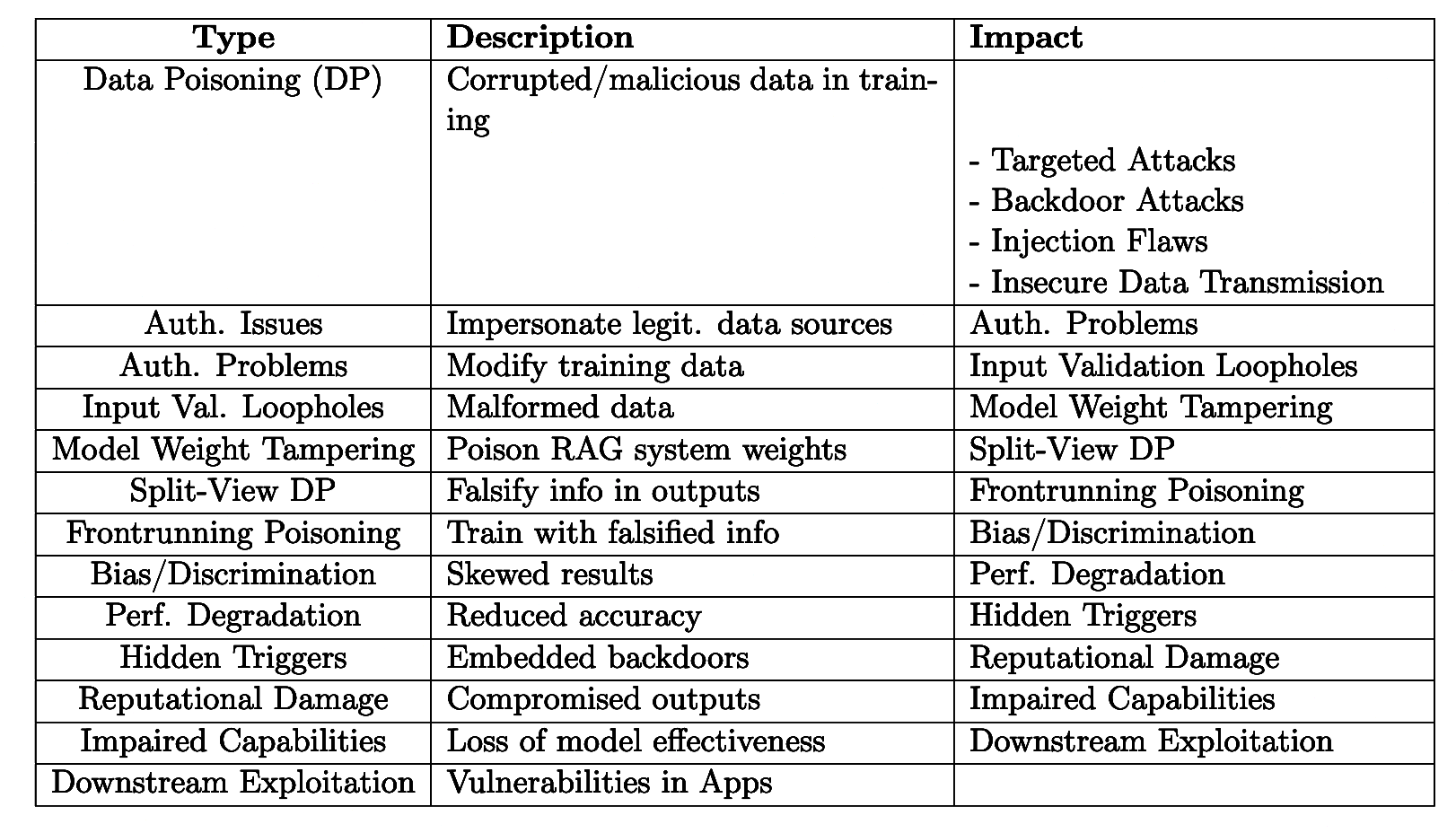} 
\caption{Data poisoning.}
\label{fig:Data_poisoning_Finance}
\end{figure}

Table \ref{table:v02} examine the possible risks (Figure \ref{fig:Data_poisoning_Finance} )\footnote{\href{https://owaspai.org/docs/ai_security_overview/}{AI Security}} \footnote{\href{https://cloudsecurityalliance.org/blog/2023/11/22/mitigating-security-risks-in-retrieval-augmented-generation-rag-llm-applications}{Risks}} \footnote{\href{https://www.nasdaq.com/press-release/elastic-security-labs-releases-guidance-to-avoid-llm-risks-and-abuses-2024-05-06}{LLM Risks}}  \cite{jiang2024turning}, associated with the application of “large language models” in the fields sentiment\footnote{\href{https://towardsdatascience.com/nlp-in-the-stock-market-8760d062eb92}{NLP in the Stock Market}} analysis, speech analysis \footnote{\href{https://www2.deloitte.com/us/en/insights/focus/cognitive-technologies/natural-language-processing-examples-in-government-data.html}{speech analysis}}\cite{shu2023llasm}, \cite{hao2023boosting}, \cite{zhang2023speechgpt},\cite{zhan2024anygpt},  \cite{dighe2024leveraging},\cite{hu2024wavllm},\cite{fang2024llama}, \cite{chen2024bestow}, etc., to natural language processing \cite{bommarito2021lexnlp},\cite{sawicki2023state},\cite{vamvourellis2022learning},\cite{ho2022natural},\cite{cheng2024adapting},\cite{trozze2024large}, for example in the documents of the Security and Exchange \footnote{\href{https://money.usnews.com/investing/term/securities-and-exchange-commission-sec}{SEC: Definition}} \footnote{\href{https://en.wikipedia.org/wiki/U.S._Securities_and_Exchange_Commission}{Securities and Exchange Commission}} \footnote{\href{https://www.sec.gov/newsroom/press-releases/2023-140}{Securities and Exchange}} Commission (SEC) \cite{fang2024llm},\cite{nie2024survey}, with regard to backdoor attacks \cite{barrett2023identifying},\cite{islam2023comprehensive},\cite{addad2024homeopathic},\cite{zou2024poisonedrag},\cite{cheng2024trojanrag}, on language models and transformers\footnote{\href{https://julsimon.medium.com/analyzing-sec-filings-with-transformers-for-fun-and-profit-505b42cef643}{Analyzing SEC filings with Transformers}} \footnote{\href{https://www.youtube.com/watch?v=GfjUJ1TnI-o}{Finance LLM for FREE with SEC Data}} \footnote{\href{https://medium.com/@getbind.co/comparing-chatgpt-bard-llama-and-custom-llm-application-for-financial-information-from-sec-e9911e612ff5}{Finance LLM for Vector database and RAG}}.

\bibliographystyle{IEEEtran}
\bibliography{IEEEabrv,Bibliography}


\EOD

\end{document}